\documentclass[preprint]{elsarticle} 

\usepackage{xcolor}
\usepackage{soul}

\usepackage[utf8]{inputenc}
\usepackage{amsmath}
\usepackage{mathtools}
\usepackage{mathalfa}
\usepackage{mathrsfs}
\usepackage{amssymb}
\usepackage{dsfont}
\usepackage{graphicx}
\usepackage{float}
\usepackage{physics}
\usepackage{amsfonts}
\usepackage{amsthm}
\usepackage{siunitx}
\usepackage{microtype}
\usepackage{cancel}
\usepackage{bm}
\usepackage[hidelinks=true, hyperfootnotes=true, bookmarks=true]{hyperref}
\usepackage{mathalfa}
\usepackage{mathrsfs}
\usepackage{verbatim}
\usepackage{import}
\usepackage[capitalise, nameinlink]{cleveref}
\usepackage{subfigure}
\usepackage{caption}

\biboptions{authoryear}
\hypersetup{colorlinks={true},linkcolor={red},citecolor=blue}
\newcommand{\dif}{\mathrm{d}}

\newcommand{\avg}[1]{\left<#1\right>}
\newcommand{\parentesi}[1]{\left(#1\right)}
\newcommand{\claudator}[1]{\left[#1\right]}
\newcommand{\limitss}[2]{_{#1}^{#2}}

\usepackage[a4paper, left=30mm, right=30mm, top=30mm, bottom=30mm]{geometry}

\journal{ } 

\begin{document}

\begin{frontmatter}

\title{Vector-borne diseases with non-stationary vector populations: the case of growing and decaying populations}


\author[1]{Àlex Giménez-Romero}

\author[1,2]{Rosa Flaquer-Galmés}

\author[1]{Manuel A. Matias\corref{cor}}
\ead{manuel@ifisc.uib-csic.es}

\cortext[cor]{Corresponding author}

\address[1]{IFISC (CSIC-UIB), Instituto de Física Interdisciplinar y Sistemas Complejos, Campus UIB\\ E-07122 Palma de Mallorca, Spain}
\address[2]{Grup de Física Estadística, Departament de Física. Facultat de Ciències, Universitat Autònoma de Barcelona, 08193 Bellaterra (Barcelona), Spain}

\begin{abstract}

Since the last century, deterministic compartmental models have emerged as powerful tools to predict and control epidemic outbreaks, in many cases helping to mitigate their impacts. A key quantity for these models is the so-called \textit{Basic Reproduction Number}, $R_0$, that measures the number of secondary infections produced by an initial infected individual in a fully susceptible population. Some methods have been developed to allow the direct computation of this quantity provided that some conditions are fulfilled, such that the model has a pre-pandemic disease-free equilibrium state. This condition is only fulfilled when the populations are \textit{stationary}. In the case of vector-borne diseases, this implies that the vector birth and death rates need to be balanced, what is not fulfilled in many realistic cases in which the vector population grow or decrease. Here we develop a vector-borne epidemic model with growing and decaying vector populations and study the conditions under which the standard methods to compute $R_0$ work and discuss an alternative when they fail. We also show that growing vector populations produce a delay in the epidemic dynamics when compared to the case of the stationary vector population. Finally, we discuss the conditions under which the model can be reduced to the SIR model with fewer compartments and parameters, which helps in solving the problem of parameter unidentifiability of many vector-borne epidemic models.

\end{abstract}


\end{frontmatter}

\section{Introduction}\label{sec:intro}
  
    Vector-borne diseases are caused by infectious agents transmitted by living organisms, called vectors,  frequently arthropods. These diseases represent a significant threat to global human health \citep{Athni_2020}, causing diseases such as malaria, dengue, yellow fever, Zika, trypanosomiasis and leishmaniasis \citep{SCHUMACHER2018352}. Vector-borne human diseases are responsible of more than 17\% of all human infectious diseases, causing millions of cases and more than $700\,000$ deaths annually \citep{WHO}. Moreover, crop production and farm profitability are also affected by bacterial \citep{HUANG20201379} and virus \citep{Bragard2013} vector-borne diseases. Some examples are the Pierce's Disease of grapevines, that has resulted in an annual cost of approximately \$100 million in California alone \citep{tumber2014pierce}, the olive quick decline syndrome, which could cause about \$5 and 17 billion of loss in Italy and Spain over the next 50 years in the absence of disease control measures \citep{Schneider2020} and the multiple diseases caused by viruses \citep{Rybicki2015}, with diseases like the tobacco mosaic, tomato spotted wilt, etc. and transmitted by aphids and other vectors.
    
    Compartmental deterministic models, e.g. the well known SIR model \citep{Kermack1927}, have been widely used in the modelling of vector-borne diseases after the work of Ross and Macdonald \citep{Macdonald1957}, that, opened the way to controlling malaria outbreaks by acting on the vectors of the disease (the \textit{Anopheles} mosquito). These models consider that both host and vector populations can be divided into different compartments describing different states of the individuals, such as susceptible, infected or dead \citep{Brauer2008}, and the time-evolution of these compartments is expressed as a system of ordinary differential equations, defining a dynamical system. Compartmental models provide a mean-field description, that imply well-mixed (in practice spatially homogeneous) populations. The well mixed approximation will be valid whenever the mean distance among hosts is smaller than the mixing length of vectors before they die. In the case of vector-borne diseases it is also equivalent to every vector effectively interacting with all the hosts and every host with all the vectors. A mean-field description is not always valid in spatially extended systems, but still it is normally the first step before writing a spatially explicit description. 
    
    The most relevant piece of information about a disease is whether an epidemic outbreak will take place. The \textit{basic reproduction number}, $R_0$, measures the number of secondary infections caused by an initial infected individual in a fully susceptible population, defining the epidemic threshold \citep{Anderson1991, VandenDriessche2017}, that determines the emergence (or not) of an outbreak. If $R_0>1$ an epidemic outbreak will occur, while there will be no outbreak otherwise. The standard way of determining $R_0$ in deterministic compartmental models assumes the existence of an initial disease-free (pre-pandemic) equilibrium, represented by the absence of infected hosts and vectors \citep{Lauko2006, Kamgang2008}. Some standard methods based on the linear stability condition of this equilibrium have been developed to allow the direct computation of $R_0$, such as the Next-Generation Matrix (NGM) method \citep{Diekmann2010}. 
    
    In the case of vector-borne diseases models typically assume that populations (both hosts and vectors) do not change with time (see e.g. \citep{Macdonald1957, Brauer2016}), assuming equal birth and death rates. This guarantees the existence of a disease-free equilibrium and the proper use of standard methods to determine $R_0$. However, this assumption could be far from reality in several pathosystems. In particular, the interaction between temperature, precipitation variations and other factors may lead to strong variations in the vector population \citep{garms1979studies,Rocklov2020}, implying that the pre-pandemic state may not be an equilibrium state and that standard methods cannot be applied.
    
    Some authors have explicitly considered more general cases in which the demographic rates are not identical or even time-dependent with a given periodicity. In the case of unbalanced birth and death rates one obtains an asymptotic stationary vector population, but not an initial disease-free equilibrium, except in the case that the initial vector population coincides with the asymptotic value. Nevertheless, the basic reproduction number of these models is often computed by means of the standard methods from the asymptotic state of the population \citep{Martcheva2008, Lashari2011, Shah2013, Zhao2020, Esteva1998}. The use of these methods is supported by the fact that the asymptotic dynamics of the system under study can be described by the subsystem in which the vector population is in its stationary value \citep{Thieme1992,Thieme1995}. Indeed in \citep{Esteva1998} it is explicitly pointed out that ``it is enough to study the asymptotic behaviour of the model''. In the case of periodic demographic rates it has been shown that the time-averaged basic reproduction number defines the epidemic threshold under some circumstances \citep{Wesley2009} and even a generalisation of the NGM method \citep{Diekmann2000} has been developed for these cases \citep{Bacaer2006, Bacaer2007}
    
    Compartmental models of vector-borne diseases have another feature that may hinder their practical applicability. Namely the fact that these models have many compartments, having to describe both hosts and vectors, and as a consequence a relatively large number of parameters. This may lead to an issue known as \textit{parameter identifiability and uncertainty} \citep{Chowel2017}, depending on the available data, that is more likely to be found in models with many compartments and parameters \citep{Roosa2019}. Usually, parameter estimation procedures are needed to connect the models with disease data, mainly using incidence or prevalence over time in the host population. Unfortunately, under many circumstances the underlying model parameters are unidentifiable, so that many different sets of parameter values produce the same model fit \citep{Kao2018}. Moreover, these parameters can be really difficult to determine from the available experimental data. Nevertheless, in some cases, mathematical manipulations can be performed to reduce the model complexity using exact or approximate relations \citep{Gimenez2021}. In such cases, the number of parameters of the models can be usually reduced in terms of new parameters defined as combinations of some of the original parameters. 
    
    The plan of this paper is as follows. In \cref{sec:model} we develop a compartmental model of vector-borne transmitted diseases with constant, but different, birth and death rates for the vectors, which allow us to describe growing and decaying vector populations. For simplicity, the model assumes that there is no horizontal (host to host) direct transmission, which is also a realistic assumption in some cases, like plant diseases. \cref{sec:results} contains the main results of the study. In particular, we show how and when traditional methods fail to estimate the $R_0$ of the model and provide an alternative way to compute it. It turns out that the validity of the standard methods depends, among other things, on some time-scales of the model. Furthermore, we discuss and apply some approximations that allow to reduce the model in favour of simpler ones, with both fewer compartments and fewer parameters. In particular, we show that if some of the parameters fulfil certain conditions, it is possible to reduce the original model with five compartments and four parameters to a SIR model, with three compartments and two parameters. It is expected that model reductions like this one significantly help in solving possible problems of parameter unidentifiability that plague these models. It is interesting to note that a model in which hosts do not interact directly, but only through vectors, in a certain limit becomes described as if hosts would infect one to each other, what is assumed in some studies without suitable confirmation. Finally, the main concluding remarks of the study are presented in \cref{sec:conclusions}.
    
\section{The model}\label{sec:model}
    
    The compartment model for vector-borne diseases that we will use to illustrate the points to be discussed in this study consists of $5$ compartments, $3$ of which describe the host population (susceptible, $S_H$, infected, $I_H$, and removed, $R_H$), while the other $2$ describe the vector population (susceptible $S_V$ and infected vectors, $I_V$). Thus, we consider that the pathogen affects only the hosts and do not consider exposed compartments. In addition, no horizontal (direct host to host) or vertical (mother to offspring for vectors) transmission is assumed. The model could be also generalised to include an exposed host compartment and the above mentioned transmission modes, which would hinder the theoretical analysis without altering the qualitative conclusions of the study. Anyhow, the absence of horizontal transmission would make the model adequate to study vector-borne diseases in which host-to-host infection is rare, like malaria, and most phytopathologies, although not for, say, zika. Finally, we do not consider host recruitment neither infected hosts becoming susceptible again.
    
\begin{figure}[H]
        \centering
        \includegraphics[width=0.7\textwidth]{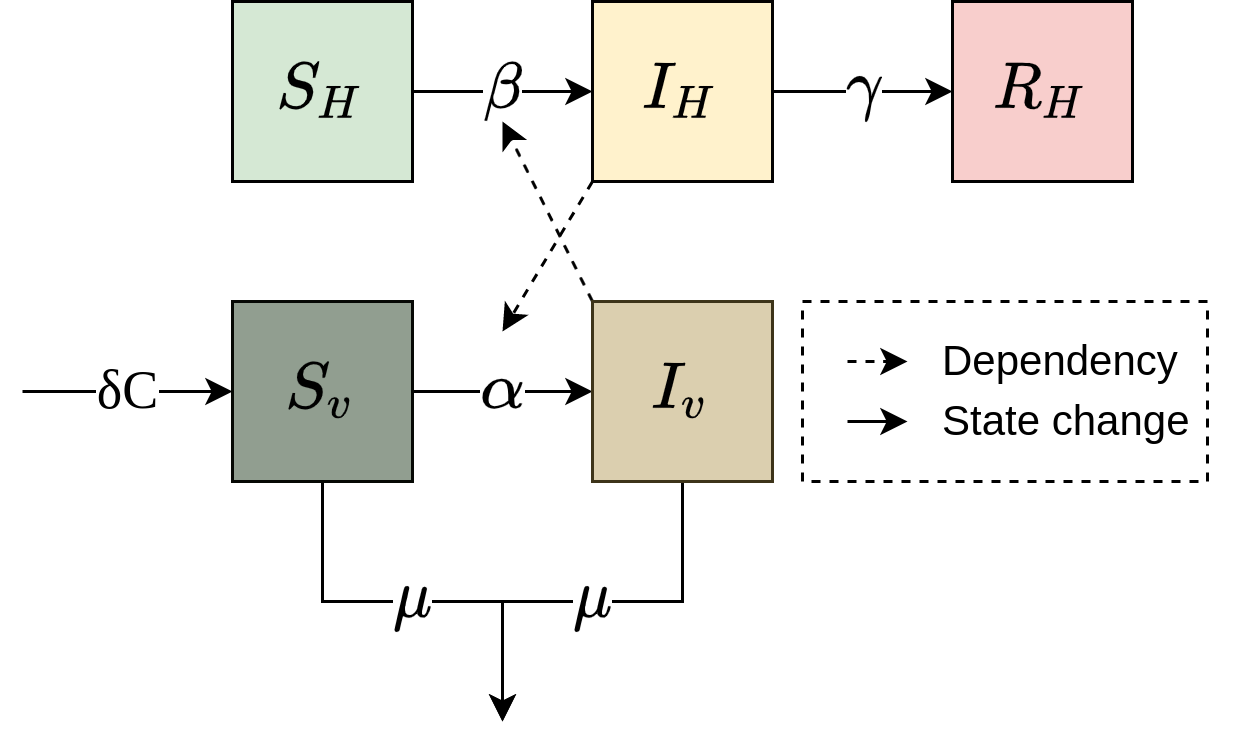}
        \caption{Schematic representation of the model n \cref{eq:SIR_v}. Boxes are the compartments in which the population is divided, solid arrows represent changes in state (so transitions between compartments), and dashed arrows depict the crossed interaction between hosts and vectors.}
        \label{fig:model_diagram}
    \end{figure}

    The model is defined according to the following processes,
    \begin{equation}\label{eq:scheme_infection}
        S_H+I_V \stackrel{\beta}{\rightarrow} I_H + I_V \quad I_H  \stackrel{\gamma}{\rightarrow} R_H \quad S_V+I_H \stackrel{\alpha}{\rightarrow} I_V+I_H \quad S_V \stackrel{\mu}{\rightarrow} \varnothing \quad I_V \stackrel{\mu}{\rightarrow} \varnothing
        \ ,
    \end{equation}   
    which are graphically described in \cref{fig:model_diagram}, being the birth of new susceptible vectors described as a source term.
    Thus, the host-vector compartmental model is written as,
    \begin{equation}\label{eq:SIR_v}
        \begin{aligned}
            \dot{S}_H &=-\beta S_H I_v / N_H \\
            \dot{I}_H &=\beta S_H I_v / N_H - \gamma I_H \\
            \dot{R}_H &=\gamma I_H \\
            \dot{S}_v &= \delta C-\alpha S_v I_H / N_H - \mu S_v \\
            \dot{I}_v &=\alpha S_v I_H / N_H - \mu I_v \ ,
        \end{aligned}
    \end{equation}
    where a standard incidence \citep{MartchevaBook} has been considered.
    
    The model describes infection of susceptible hosts, $S_H$, at a rate $\beta$ through their interaction with infected vectors, $I_v$, while susceptible vectors, $S_v$, are infected at a rate $\alpha$ through their interaction with infected hosts $I_H$. Infected hosts exit the infected compartment at rate $\gamma$, while infected vectors stay infected the rest of their lifetime, as we consider that the pathogen does not affect them, as it is customary. Vectors die naturally (or disappear from the population by some mechanism) at rate $\mu$ and are born (appear) at a constant rate $\delta$ being susceptible. The constant term $C$ sets the scale of the stationary value of the vector population. \cref{fig:model_diagram} shows an schematic representation of the model and we refer to \citep{Brauer2016} for a similar model of vector-borne diseases. The mentioned model includes exposed compartments and horizontal transmission, but assumes that the birth and death rate of vectors are identical, and thus, the population does not change with time and stays as fixed by the initial condition.
    
\subsection{Preliminary analysis of the model} \label{sec:Prelimanalysis}

    From \cref{eq:SIR_v} it is straightforward to verify that the population of hosts remains constant over time, $N_H=S_H+I_H+R_H$, while the vector population fulfils,
    \begin{equation}\label{eq:dif_eq_Nv}
        \dot{N}_v=\dot{S}_v+\dot{I}_v=-\mu\parentesi{S_v+I_v}+\delta C=-\mu N_v + \delta C \ ,
    \end{equation}
    which can be solved to yield,
    \begin{equation}\label{eq:Nv_t}
        N_v(t)=\frac{\delta}{\mu}C + \parentesi{N_v(0)-\frac{\delta}{\mu}C}e^{-\mu t} \ .
    \end{equation}
    From \cref{eq:Nv_t} the stationary value for the vector population, $N_v^*$, can be computed as,
    \begin{equation}
        N_v^*=\lim_{t\to\infty}N_v(t)=\frac{\delta}{\mu}C \ .
        \label{eq:asympt}
    \end{equation}
    Thus, if the initial population of vectors is below (above) the stationary value, the vector population will grow (decrease) until it reaches the stationary value. On the other hand, if $N_v(0)=N_v^*=\delta C/\mu$ the initial population of vectors is already at the stationary state. The initial condition for the vector population can be written in terms of its stationary value \cref{eq:asympt}, $N_v(0)=fN_v^*$, where both $f<1$ and $f>1$ are possible, so that one gets,
    \begin{equation}\label{eq:Nv_t_fraction}
        N_v(t)=N_v^*\claudator{1+\parentesi{f-1}e^{-\mu t})} \ .
    \end{equation}
    
    We note that vector-borne disease models that assume constant vector populations (e.g.\citep{Brauer2016}) can be recovered by setting $\delta=\mu$ and $C=N_v(0)$, so that any initial condition for the vector population is stationary, i.e. $\dot{N}_v=0$ in \cref{eq:dif_eq_Nv} and $N_v(t)=N_v(0)$. 
    
\section{Results} \label{sec:results}

\subsection{The effect of non-stationary vector populations into the epidemic threshold and disease dynamics}
\label{sec:R0statvp}

    Let us start with the case in which the birth $\delta$ and death $\mu$ vector rates are identical and the initial condition of the vector population is already at its stationary value, $N_v(0)=N_v^*$, case in which the total vector population remains unchanged, as discussed in \cref{sec:Prelimanalysis}.
    In such a case, the initial disease free state of the model, given by $I_H(0)=I_v(0)=0$, is a fixed point (equilibrium state) of the dynamical system \cref{eq:SIR_v} independently of the other initial conditions for the host and vector populations. This allows the definition of the basic reproduction number, $R_0$, using standard methods such as linear stability analysis or the Next-Generation Matrix (NGM) method \citep{Diekmann2010} (see \ref{app:R0_standar_methods}).
    
    The situation is different when $\delta\neq\mu$, as then the total vector population will vary with time if the initial condition, $N_v(0)$, is not identical to the asymptotic value at large times, $N_v^*$, and thus an initial disease-free state is not an equilibrium (fixed point) of the model. However, in the literature it is customary to apply the standard techniques, i.e. NGM, to compute $R_0$ using the vector population in the asymptotic state, that is the post-pandemic disease-free equilibrium \citep{Martcheva2008, Lashari2011, Shah2013, Zhao2020, Esteva1998}.
    The use of these methods is supported by the fact that the asymptotic dynamics of the model converges to the dynamics of the subsystem where the vector population is stationary \citep{Thieme1992,Thieme1995}. In both cases the basic reproduction number is given by,
    \begin{equation}\label{eq:R0_asympt}
       R_0=\frac{\beta\alpha}{\mu\gamma}\frac{S_H(0)}{N_H}N_v^* \ .
    \end{equation}
    As usual, $R_0$ accounts for the number of secondary infections produced by an infected individual in one generation and controls the threshold behaviour of the model: for $R_0<1$ the epidemic dies out and for $R_0>1$ an outbreak occurs. By one generation we refer to the typical time in which new infections can be produced, being the generation time in our model,
    \begin{equation}
    t_g=1/\gamma + 1/\mu\ .
    \label{eq:generationtime}
    \end{equation}
    
            \begin{figure}
        \centering
        \includegraphics[width=\textwidth]{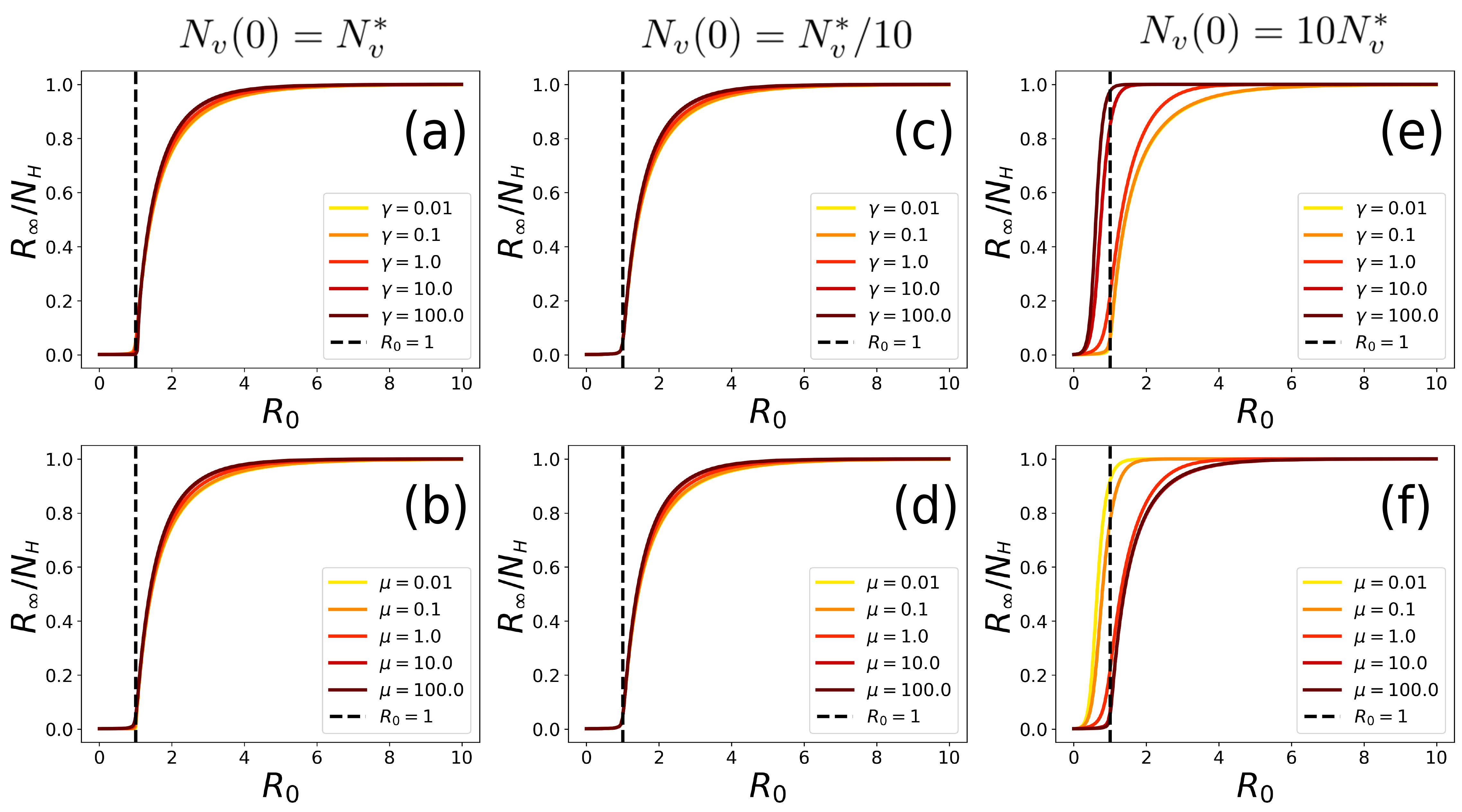}
        \caption{Numerical verification of the predictive  power of the basic reproduction number relation \cref{eq:R0_asympt}, by plotting the final size of the epidemic, $R(\infty)/N_H$ as function of $R_0$. In panels (a),(b) the initial vector population is in the stationary value, in panels (c),(d) is below, $N_V^*/10$, and in panels (e),(f) above, $10 N_V^*$. Panels (a),(c),(e) show realisations for different $\gamma$ values with a fixed $\mu=1$ baseline value. Panels (b),(d),(f) show realisations for different $\mu$ values with a fixed $\gamma=1$ baseline value.}
        \label{fig:R0_check_stationary}
    \end{figure}
    
    Here we show that this result can only be used in the more general case provided that some conditions are fulfilled, but not \textit{always}. Let us first illustrate the different regimes in which \cref{eq:R0_asympt} is predictive or not about the onset of the epidemic. In \cref{fig:R0_check_stationary} the final size of the epidemic, $R_{\infty}/N_H$, is plotted as a function of $R_0$. \cref{fig:R0_check_stationary}(a)-(d) show that \cref{eq:R0_asympt} does indeed regulate the onset of an epidemic when the initial vector population is in its stationary value or below it. This result is general and does not depend on the time-scales of the system, $1/\gamma$ and $1/\mu$, and so all curves in these panels behave similarly. In contrast, \cref{fig:R0_check_stationary}(e)-(f) shows how the $R_0$ calculated according to \cref{eq:R0_asympt} is not predictive of the onset of epidemic outbreak, this happening when the initial vector population is larger than the stationary value. Thus, for $R_0<1$ (computed using  \cref{eq:R0_asympt}) severe outbreaks appear, yielding mortalities even above 80\% of the total population. However, one can observe that  as $\mu$ is increased, or $\gamma$ decreased, the predictive power of \cref{eq:R0_asympt} is progressively recovered.
    
    The observed behaviour and the regimes of validity of \cref{eq:R0_asympt} can be explained readily. In essence, only if the vector population reaches its stationary value before infected hosts have produced new infections the onset of an epidemic can be characterised by \cref{eq:R0_asympt}, but not otherwise. Given these different behaviours, let us discuss separately the cases $f>1$ and $f<1$, with $N_v(0)=fN_v^*$, namely when the initial vector population is above and below its stationary value, this is, decaying and growing vector populations towards the asymptotic value.
    
    Let us start with the case of decaying vector populations, $f>1$.
    From \cref{eq:Nv_t_fraction}, the time to approach the stationary value, $t^*$, fulfils
    \begin{equation}
        \parentesi{1+\epsilon}N_v^*=N_v^*\claudator{1+(f+1)e^{-\mu t^*}} \ ,
    \end{equation}
    where $\epsilon\to 0$ is a small parameter controlling the amount by which the vector population differs from its asymptotic value at time $t^*$. Thus, the time to approach the stationary value, with precision $\epsilon$, is given by
    \begin{equation}
        t^*=-\frac{1}{\mu}\ln(\frac{\epsilon}{f-1})=\frac{1}{\mu}\abs{\ln{\frac{\epsilon}{f-1}}} \ ,
    \end{equation}
    where the last equality assumes that the small parameter $\epsilon$ satisfies $\epsilon<(f-1)$, $f>1$.
    
    If the vector population reaches its stationary value before infected hosts have had time to generate new infections then $R_0$ as determined from \cref{eq:R0_asympt} is a good prediction of the onset for an epidemic, what is equivalent to the condition that $t^*$ is much smaller than the hosts infectious period, $t^*\ll1/\gamma$,
    \begin{equation}\label{eq:timescales_condition}
        \frac{1}{\gamma} \gg \frac{1}{\mu}\abs{\ln{\frac{\epsilon}{f-1}}} \quad \textrm{or} \quad \frac{\mu}{\gamma}\gg\abs{\ln{\frac{\epsilon}{f-1}}}  \ .
    \end{equation}
    Otherwise, \cref{eq:R0_asympt} will not be predictive of the epidemic onset, and as shown in \cref{fig:R0_check_stationary}(e-f) one may have outbreaks with a substantial final size with $R_0<1$.\\
    
    In the case of growing vector populations, $f<1$, if $R_0<1$ an outbreak cannot occur at all, because $R_0$ is calculated with the asymptotic population, $N_v^*$, that is larger that the vector population at any finite time, $N_v(t)<N_v^* \ \forall t$, and so the threshold condition is never attained. In the $R_0>1$ case the behaviour will be richer, and it will depend on the initial condition, $N_v(0)$. One can define an instantaneous basic reproductive number,
    \begin{equation}\label{eq:R0i}
        R_0^{(i)}(t)=\frac{\beta\alpha}{\mu\gamma}\frac{S_H(0)}{N_H} N_v(t)=R_0\frac{N_v(t)}{N_v^*} \ ,
    \end{equation}
    using $N_v(t)$ instead of $N_v^*$, with $R_0^{(i)}(t)<R_0 \ \forall t$ because the vector population grows. In particular, if $R_0^{(i)}(0)>1$ there will be an outbreak occurring for short times, and the population of infected hosts will start growing. If instead, $R_0^i(0)<1$, and as $R_0>1$ with $R_0$ being calculated with the asymptotic state, there must be an intermediate time, say $t_D$, for which $R_0^{(i)}(t_D)=1$. Thus, from $t>t_D$ an outbreak will occur, not initially but after a finite time, that induces a delay in the outbreak, and the infected host population will start growing.
    
    The difference between the original and the delayed dynamics stems from the waiting time to reach $R_0^{(i)}=1$, $t_D$, plus the non-linear effect associated to a new initial condition for the epidemic outbreak at $t_D$. Thus, in the case that $R_0>1$ and $R_0^{(i)}(0)<1$, from \cref{eq:R0i} and \cref{eq:Nv_t_fraction} we can analytically approximate the delay as the time needed to reach $R_0^{(i)}(t_D)=1$,
    \begin{equation}
        {1+(f-1)e^{-\mu t_D}}=\frac{1}{R_0} \ ,
    \end{equation}
    which yields the relation,
    \begin{equation}\label{eq:delay}
        t_D=-\frac{1}{\mu}\ln\claudator{\frac{1-R_0}{\parentesi{f-1}R_0}}\ ,
    \end{equation}
    where the argument of the logarithm is always positive because $R_0>1$ and $f<1$. \cref{eq:delay} is only valid if $f<1/R_0$, for $R_0^{(i)}(0)=f R_0<1$, as if otherwise $R_0^{(i)}>1$ the outbreak would already occur initially.
    
    \begin{figure}[H]
        \centering
        \includegraphics[width=\textwidth]{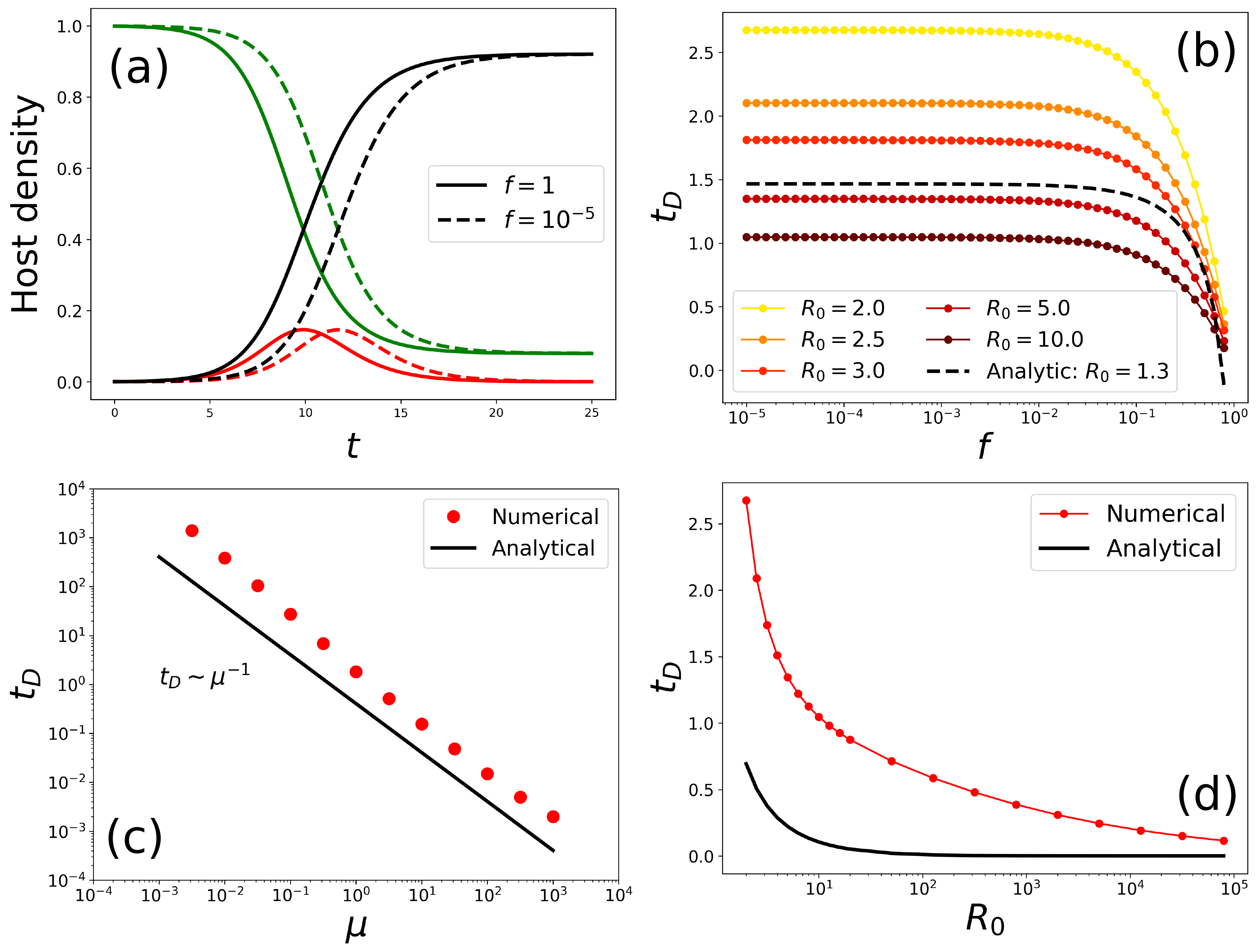}
        \caption{Numerical study of the delay induced by growing vector populations. (a) Comparison of hosts dynamics for a stationary vector population ($f=1$) and a growing vector population ($f=10^{-5}$). (b) Time delay as function of $f$ for different values of the basic reproduction number $R_0$. (c) Time delay as function of the vector natural death rate. (d) Time delay as function of the basic reproduction number, $R_0$, with $f=10^{-5}$.}
        \label{fig:delay}
    \end{figure}
    
    From \cref{eq:delay} one can see that when the initial vector population is far enough from its stationary value, $f\rightarrow 0$, the delay saturates to a constant value, instead of increasing. This is,
    \begin{equation}
        \lim_{f\to0}t_D= -\frac{1}{\mu}\ln(\frac{R_0-1}{R_0})=\frac{1}{\mu}\ln(\frac{R_0}{R_0-1}) \ .
        \label{eq:limitd}
    \end{equation}
    In addition, for increasing values of the basic reproduction number, $R_0$, the delay tends to vanish, and from \cref{eq:limitd}. This is,
    \begin{equation}
    \label{eq:limit_tD_infty}
       \lim_{R_0\to\infty}t_D=\frac{1}{\mu}\ln(1)=0\ ,
    \end{equation}
    where the limit $f\rightarrow 0$ is taken simultaneously to guarantee that $R_0^{(i)}(0)=f R_0<1$. On the other hand the delay, $t_D$, scales with the vectors lifetime,
    \begin{equation}
        t_D\sim\frac{1}{\mu}=\tau_v \ .
    \end{equation}
    
    \cref{fig:delay}(a) shows an example of the time delay caused in the hosts dynamics when the vector population grows from an initial condition far from the stationary value. In \cref{fig:delay}(b) we can qualitatively observe that all the predicted properties of the delay are fulfilled, namely, the time delay saturates for low $f$ values and decreases with increasing $R_0$. Although the analytical expression (black dashed line) is clearly not exact due to nonlinear effects, \cref{eq:delay} captures the basic trends of the time delay, $t_D$. This is clear from \cref{fig:delay}(c), that shows that the delay scales with $1/\mu$ and in \cref{fig:delay}(d) that shows that the delay tends to $0$ in the limit $R_0\rightarrow\infty$, in agreement with the prediction of \cref{eq:limit_tD_infty}.
    
\subsection{The basic reproduction number for non-stationary vector populations}

    As shown in the previous section, traditional methods to compute the basic reproduction number fail in the case of epidemic models with decaying vector populations, $f>1$, unless the time scale of vector population fulfils several conditions, as illustrated in \cref{sec:R0statvp}. Here we introduce an effective, average definition of $R_0$, useful to predict the epidemic onset for vector-borne diseases with decaying vector populations, i.e. the case where traditional methods fail. It is defined as the \textit{average} number of infections produced by an infected individual in \textit{one generation} \cref{eq:generationtime}, 
    \begin{equation}
        \overline{R_0}=\avg{R_{0}^{i}(t)}\Big|\limitss{0}{t_g}=R_0\claudator{1-\frac{1}{\tau}\parentesi{f-1}\parentesi{e^{-\tau}-1}}=R_0\cdot\mathcal{F}
       \label{eq:R0_non_stationary}
    \end{equation}
    where $\tau=1+\mu/\gamma$ and $\mathcal{F}$ is the expression in brackets accounting for the effect of the decaying vector population on the stationary $R_0$ (see \ref{app:R0_non_stationary} for the full derivation of \cref{eq:R0_non_stationary}).
     
         \begin{figure}[H]
        \centering
        \includegraphics[width=0.9\textwidth]{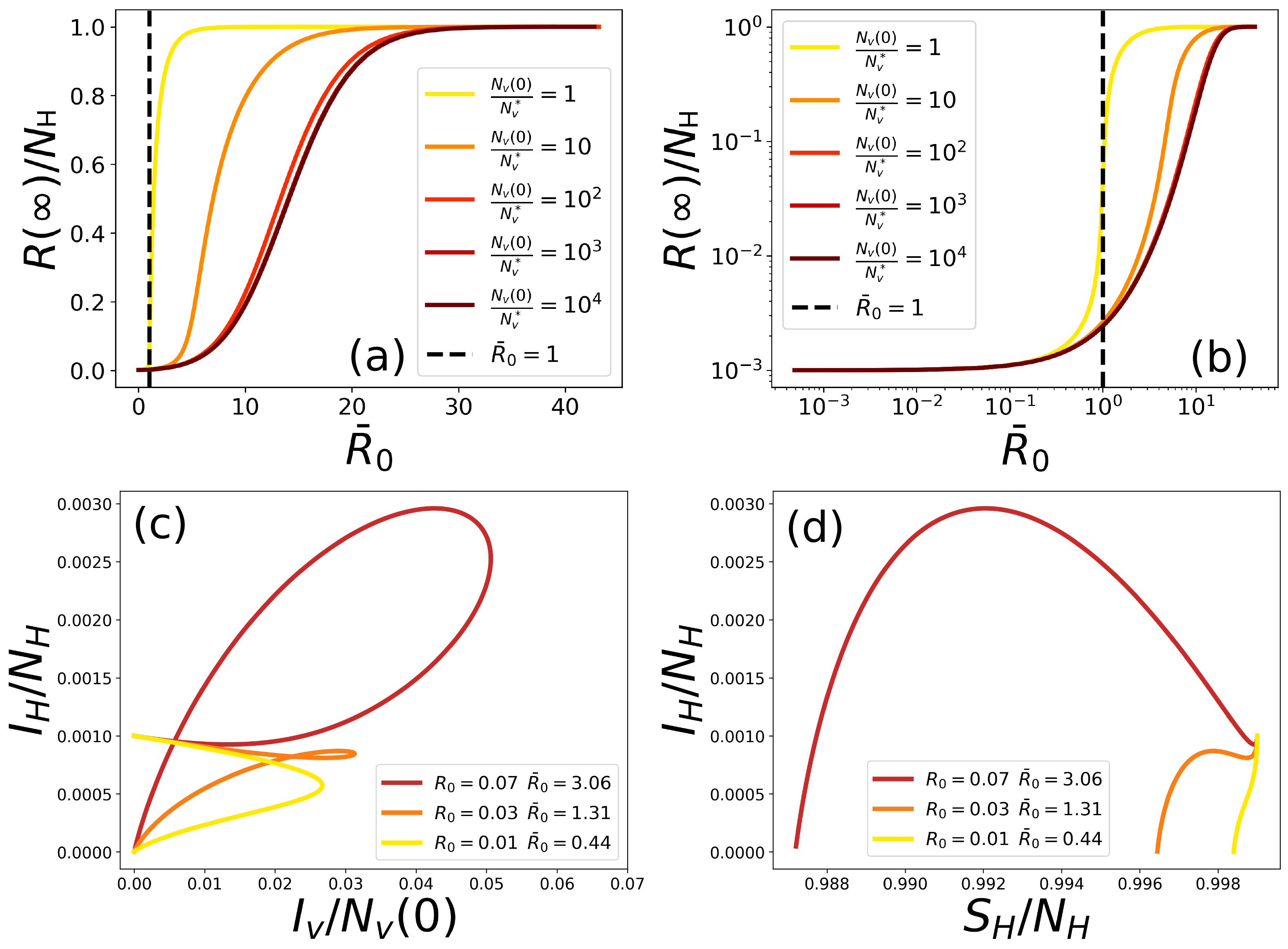}
        \caption{Numerical verification of the expression for the basic reproduction number for vector-borne diseases with decaying vector populations \cref{eq:R0_non_stationary}. Final size of the epidemic as a function of the basic reproduction number in panels: (a) linear scale; (b) logarithmic scale. Phase space trajectories in panels: (c) $I_H/N_H$ vs $I_v/N_v(0)$ and (d) $I_H/N_H$ vs $S_H/N_H$, where an initial condition $I_H(0)/N_H=0.01, S_H(0)/N_H=0.99$ and $I_v(0)/N_V(0)=0$ has been used for the $3$ cases. $\mu=\gamma$ has been used in all the simulations.}
        \label{fig:R0_check_mean_value}
    \end{figure}
    
    A first observation is that $\overline{R_0}>R_0$ always. This stems from the fact that $\tau>1$, so that $e^{-\tau}-1<0$, and $f-1>0$, which yields $\mathcal{F}>1$. This discussion unravels why standard methods fail to predict the onset of an epidemic under decaying vector populations. Another important point is that if $\mu/\gamma\gg 1$, which implies $\tau\gg1$, then $\overline{R_0}\to R_0$ as,
    \begin{equation}
        \lim_{\tau\gg1}\mathcal{F} =\lim_{\tau\gg1}\claudator{1-\frac{1}{\tau}\parentesi{f-1}\parentesi{e^{-\tau}-1}}=1+\frac{f-1}{\tau} \ ,
    \end{equation}
    and if furthermore $\tau\sim\frac{\mu}{\gamma}\gg (f-1)$ then $\mathcal{F}\to 1$ and $\overline{R_0}\to R_0$. This is in agreement with the discussion in \cref{sec:R0statvp} showing that the $R_0$ computed from standard methods works if $\mu\gg\gamma$.
    
    \cref{fig:R0_check_mean_value}(a-b) contrasts numerically the validity of \cref{eq:R0_non_stationary} to predict the final size of the epidemic as a function of the general basic reproduction number, $\overline{R_0}$, in linear and logarithmic scale, respectively. We observe that, independently of the initial condition of vectors, the outbreak occurs for $\overline{R_0}>1$. However, we may notice that for large values of the initial condition of vectors the final size of the epidemic grows more slowly, so that larger values of $\overline{R_0}$ are needed to produce a proper outbreak. This can be explained by the fact that for $\overline{R_0}$ slightly above the threshold, $\overline{R_0}=1$, and large values of $f=N_v(0)/N_v^*$, infections are produced only in the transient period of the dynamics, as $R_0<1$. This is, while the vector population is decaying to its stationary value, the vectors are able to produce new infections, but once the vector population reaches the stationary value, the epidemics stops. This transmission mechanism is radically different to that of vector-borne diseases with stationary vector populations in which the pre-pandemic disease-free state is an equilibrium of the system. The phase-space plots in \cref{fig:R0_check_mean_value}(c-d) show that the time-averaged basic reproduction number $\overline{R_0}$ is able to accurately predict the conditions under which the infected host population will grow, in contrast with $R_0$ computed in the post-pandemic fixed point. In essence, for $\overline{R_0}>1$ the infected host population, $I_H$, grows before reaching the absorbing state, $I_H=I_v=0$, while for $\overline{R_0}<1$ the infected host population is monotonically decreasing. We note that \cref{eq:R0_non_stationary} is similar to the time-averaged basic reproduction number presented in \citep{Wesley2009} for the periodic case, which is a first-order approximation to the \textit{true} basic reproductive number \citep{Bacaer2006}.
    
\subsection{Fast-slow approximation}
    
    The original $5$-D \cref{eq:SIR_v} model is certainly not amenable to mathematical analyses due to its high phase-space dimensionality and the fact that it depends on $4$ parameters. Moreover, in a real-case application, if the parameters conforming the model are not known the model could suffer from parameter unidentifiability. However, some approximations can be performed to reduce the mathematical complexity of the model, as for instance a fast-slow (or adiabatic) approximation. 
    
    If the time-scale of the vector population evolution is faster than that of the infected hosts, what is expected to be a good approximation in many practical cases, the vector population will almost instantaneously adapt to its stationary value. Thus, if $1/\mu\ll1/\gamma$, or equivalently if $\mu\gg\gamma$, we can rewrite the time derivative of the vector infected population as
    \begin{equation}
        \epsilon\dot{I}_v=\frac{\alpha}{\mu}S_v\frac{I_H}{N_H} - I_v \ ,
    \end{equation}
    with $\epsilon=1/\mu$ being a small parameter. Then, $\dot{I_v}$ can be neglected and the infected vector population can be obtained from the relationship,
    \begin{equation}\label{eq:Iv_timescale_approx}
        I_v\approx\frac{\alpha}{\mu}\frac{S_v I_H}{N_H} \ .
    \end{equation}
    
    Substituting \cref{eq:Iv_timescale_approx} into the original system \cref{eq:SIR_v} and the identity $N_v(t)=S_v(t)+I_v(t)$ we obtain the following reduced system,
    \begin{equation}\label{eq:reduced_model}
        \begin{aligned}
            \dot{S}_H &=-\frac{\beta\alpha N_v(t)}{\mu N_H + \alpha I_H}\frac{S_HI_H}{N_H} \\
            \dot{I}_H &=\frac{\beta\alpha N_v(t)}{\mu N_H + \alpha I_H}\frac{S_HI_H}{N_H}- \gamma I_H \\
            \dot{R}_H &=\gamma I_H \ ,
        \end{aligned}
    \end{equation}
    which can be understood as a SIR model with time-dependent coefficients,
    \begin{equation}\label{eq:SIR-time_dependent_coef}
        \begin{aligned}
            \dot{S}_H &=-\xi(t)\frac{S_HI_H}{N_H} \\
            \dot{I}_H &=\xi(t)\frac{S_HI_H}{N_H}- \gamma I_H \\
            \dot{R}_H &=\gamma I_H \ .
        \end{aligned}
    \end{equation}
    A SIR model with time-dependent periodic coefficients was discussed in \citep{Bacaer2009}.
    
    Moreover, if $f\neq1$ the above mentioned timescales relationship 
    must fulfil $\displaystyle\frac{\mu}{\gamma}\gg\abs{\ln{\frac{\epsilon}{f-1}}}$ (cf. \cref{eq:timescales_condition}) and not only $\displaystyle\frac{\mu}{\gamma}\gg 1$. It is important to notice that the presence of horizontal transmission would simply rescale the coefficient $\xi(t)$, and the SIR reduction \cref{eq:SIR-time_dependent_coef} would keep its validity.
    
\begin{figure}[H]
        \centering
        \includegraphics[width=\textwidth]{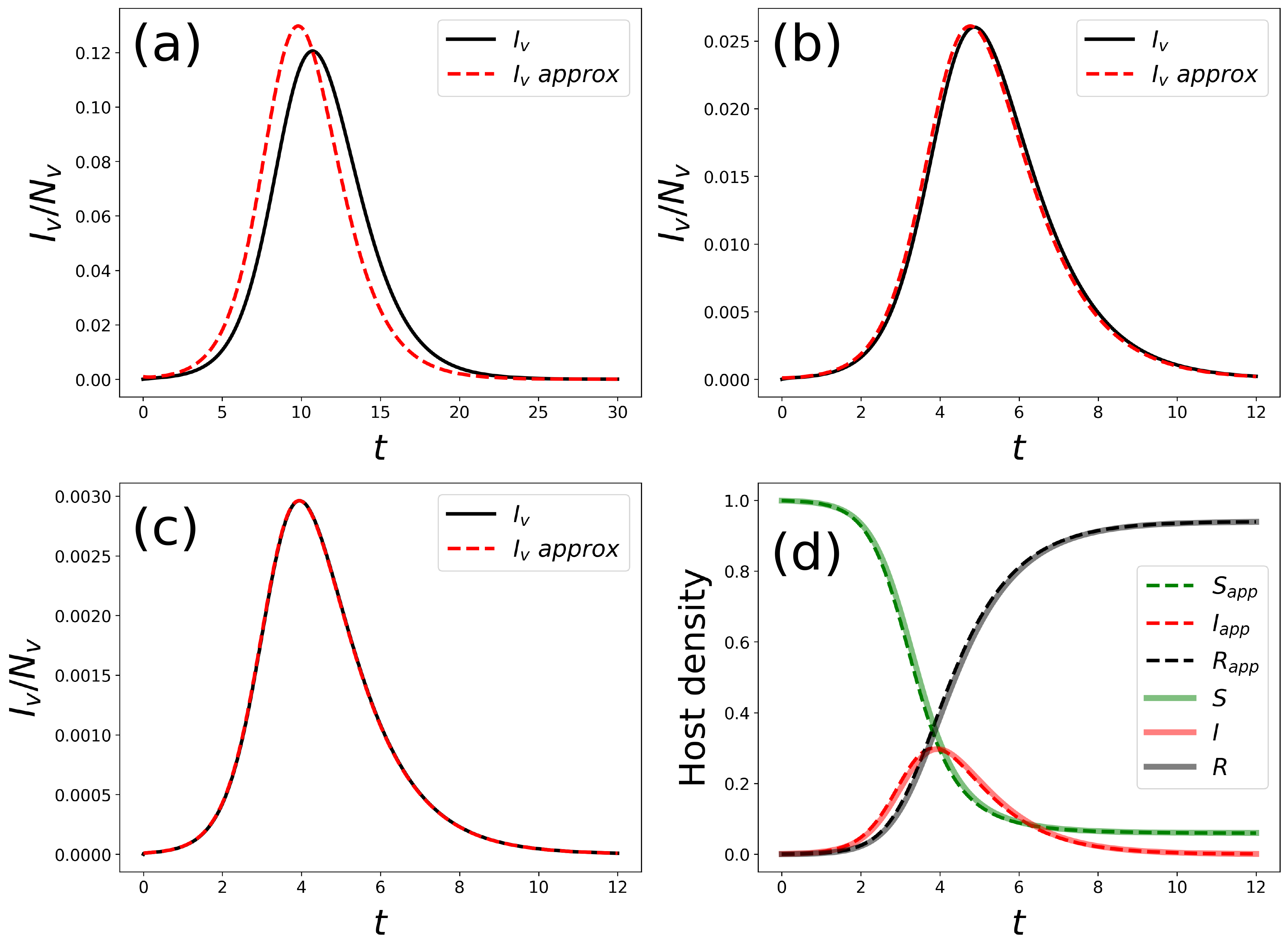}
        \caption{Numerical verification of the time-scale approximation (\cref{eq:Iv_timescale_approx}) with $N_H=100$, $\alpha=\gamma=1$. $\beta$ is chosen such that $R_0=3$. (a) $\mu=1$, (b) $\mu=10$, (c) $\mu=100$. Panel (d) shows a comparison between the approximate and original models for the parameters used in (c), where the approximated models is expected to represent well the original one.}
        \label{fig:timescale_approx}
    \end{figure}

    In \cref{fig:timescale_approx} we numerically verify the validity of the presented fast-slow approximation. As expected, we observe that the approximation breaks down for $\mu\sim\gamma$ (\cref{fig:timescale_approx}(a)), while as 
    $\mu$ becomes larger than $\gamma$ the approximation improves \cref{fig:timescale_approx}(b) and it becomes quantitative when $\mu\gg\gamma$, \cref{fig:timescale_approx}(c). Finally, we show in \cref{fig:timescale_approx}(d) a comparison between the dynamics of the hosts using both the original and the approximated model using the same parameters than in \cref{fig:timescale_approx}(c), where the results of both models are expected to converge.
    
\subsection{Reduction to a SIR model}
    
    The conditions for which the time-scale approximation is valid, $\mu\gg\gamma$, imply that the vector population will reach its stationary value almost instantaneously, so that $N_v(t)\approx N_v^*$. Thus, the system in \cref{eq:reduced_model} can be written approximately as the following SIR-like model,
    \begin{equation}\label{eq:SIR_like}
        \begin{aligned}
            \dot{S}_H &=-\beta'\frac{S_H I_H}{\lambda N_H + I_H} \\
            \dot{I}_H &=\beta'\frac{S_H I_H}{\lambda N_H + I_H}- \gamma I_H \\
            \dot{R}_H &=\gamma I_H \ ,
        \end{aligned}
    \end{equation}
    where $\beta'=\beta N_v^*/N_H$ and $\lambda=\mu/\alpha$.
    
    Furthermore, if $\lambda N_H \gg I_H$ (which is indeed plausible in this limit) \cref{eq:SIR_like} the model can be written as a SIR model with constant coefficients,
    \begin{equation}\label{eq:SIR}
        \begin{aligned}
            \dot{S}_H &=-\beta_{eff}\frac{S_HI_H}{N_H} \\
            \dot{I}_H &=\beta_{eff}\frac{S_HI_H}{N_H}- \gamma I_H \\
            \dot{R}_H &=\gamma I_H \ ,
        \end{aligned}
    \end{equation}
    where $\displaystyle\beta_{eff}=\frac{\beta'}{\lambda}=\frac{\beta\alpha N_v^*}{\mu N_H}$.

    \begin{figure}[H]
        \centering
        \includegraphics[width=\textwidth]{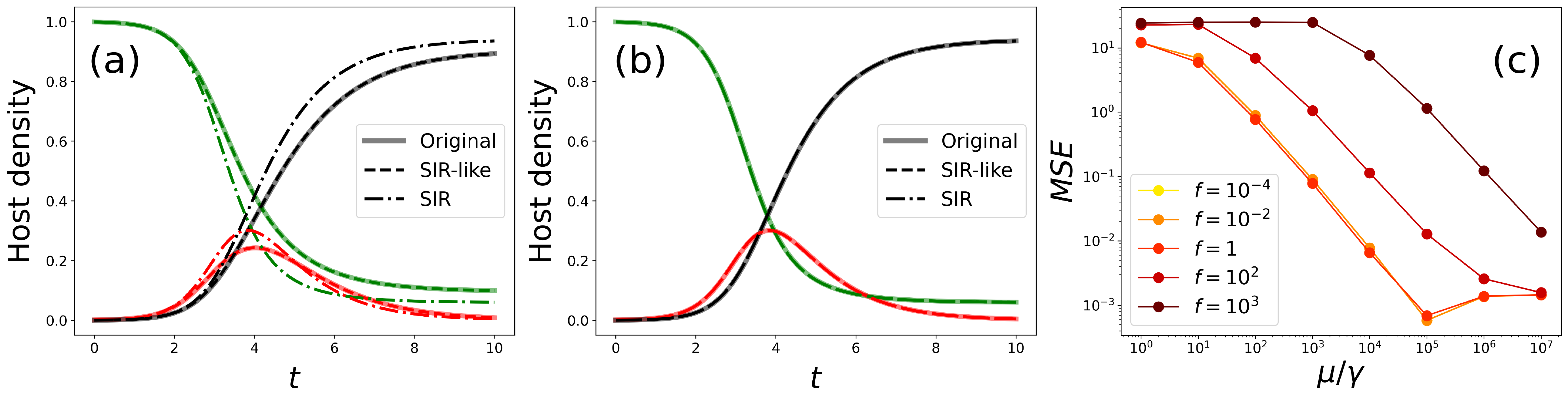}
        \caption{Comparison between the original model and the reductions, \cref{eq:SIR_like} (SIR-like) and \cref{eq:SIR} (SIR) with $N=100$, $\mu/\gamma=10^{3}$ and $f=1$. $\beta$ was chosen such that $R_0=3$.  (a) $\lambda=1$, (b) $\lambda=10^{-3}$, (c) Mean Squared Error between the original model and the SIR approximations as function of the ratio $\mu/\gamma$ and $f$.}
        \label{fig:SIR_like_approx}
    \end{figure}
    
    In \cref{fig:SIR_like_approx} we show the validity of the reduced models \cref{eq:SIR_like} and \cref{eq:SIR}. \cref{fig:SIR_like_approx}(a) shows that the SIR-like model (\cref{eq:SIR_like}) works when the time-scale approximation can be performed (as $\mu/\gamma\gg1$) but the SIR model fails when the condition $\lambda N_H \gg I_H$ is not fulfilled. Conversely, in \cref{fig:SIR_like_approx}(b) we show that as $\lambda N_H \gg I_H$ is fulfilled, then the SIR model perfectly matches the original model. Finally, \cref{fig:SIR_like_approx}(c) shows the decrease in the mean squared error of the approximation as the condition \cref{eq:timescales_condition} is fulfilled for different values of $f$.
    
\section{Conclusions}\label{sec:conclusions}
    
    In the present work we have analysed several features of a compartmental deterministic model for vector-borne diseases with $3$ compartments for hosts and $2$ for vectors, that does not consider neither horizontal nor vertical transmission. The focus is to study the behaviour of the model in the case that the vector population is not stationary. In this case, the pre-pandemic disease-free state is not a fixed point (equilibrium state) of the dynamical system, and, in principle, the methods that are customarily used to determine the basic reproduction number, $R_0$ do not work. This is so because these methods determine the onset of an outbreak by performing a linear stability analysis of the disease-free state, assuming that it is a fixed point of the model. A common assumption made in the literature is to determine $R_0$ from the asymptotic state for the vectors (if it is not an extinction state).

    We have analysed several initial conditions of the vector population, characterising different regimes. In the case that the initial condition for the number of vectors is below the asymptotic state, implying that the vector population overall grows, then $R_0$ as determined from the asymptotic state correctly predicts the existence (or not) of an epidemic outbreak, but with a temporal delay in its appearance. This result contrasts with the situation in which the initial state is above the asymptotic state, with an overall decrease in the vector population. In this case $R_0$ determined from the asymptotic state may fail badly, predicting no outbreak while a large fraction of the population might get infected. We present a simple, albeit useful, generalisation of $R_0$ that is able to give a reasonable prediction of the epidemic threshold for decaying populations, including the case in which vectors become extinct, a case in which the asymptotic estimation to determine $R_0$ cannot be applied.
    
    Compartmental models of vector-borne diseases usually have many compartments and parameters, which can lead to a problem of parameter unidentifiability. The model analysed here is not an exception, and when applied to real-world cases many different combinations of the parameters could be able to reproduce the available data. Thus, in order to facilitate the application of the model to experimental data, we have studied a useful fast-slow (or adiabatic) approximation that allows to reduce the model if the parameters fulfil certain conditions. In particular, our study shows that under quite realistic assumptions (the typical timescale of hosts infection and death is much slower than vector timescales) it is possible to obtain a reduced SIR model. We recall that this reduction implies that, under these assumptions, the process by which hosts (that could be immobile) get infected through the action of vectors is equivalent to a direct interaction among hosts.
    
    The deterministic compartmental model analysed here, with some modifications, is a clear candidate to study many vector-borne diseases, in particular phytopathologies. Furthermore, in case of parameter unidientifiability the model reductions performed in this work could be useful to solve this issue. In any case, this description is still idealised, as compartmental models imply a well-mixed assumption in which space is not explicitly described. This kind of representations are not always applicable to real-world scenarios although are useful as a first approximation. Thus, future research should focus on the integration of space and vector mobility in the model to account for more realistic situations.
    
\section*{Acknowledgments}
AGR and MAM acknowledge financial support from Grant RTI2018-095441-B-C22 (SuMaEco) funded by MCIN/AEI/10.13039/501100011033 and by “ERDF A way of making Europe" and from the María de Maeztu Program for Units of Excellence in R\&D (No. MDM-2017-0711).

\appendix

\section{Calculation of $R_0$ from standard methods}\label{app:R0_standar_methods}

    The standard methods of calculation of $R_0$ are based in the linear stability analysis of the disease-free equilibrium, either directly, through the linear analysis of the fixed point, that yields the stability condition from which $R_0$ can be obtained, or using the Next Generation Method (NGM) \citep{Diekmann2010} that provides directly $R_0$ by solving a suitable linear problem. Customarily these methods are applied to a pre-pandemic disease-free equilibrium, but as there is no such state in the case of non-stationary populations, here a similar approach is applied to a post-pandemic or asymptotic disease-free equilibrium.

    \subsection*{Linear stability analysis}
    
    In order to perform the linear stability analysis of the fixed point ($I_H=I_v=0$) we first need to compute the Jacobian matrix, $J$,
    \begin{equation}
			J = \begin{pmatrix}
				-\beta \frac{I_v}{N_H} & 0 & 0 & - \beta \frac{S_H}{N_H} \\
				\beta  \frac{I_v}{N_H}  & -\gamma & 0 & \beta \frac{S_H}{N_H}   \\
				0 & -\alpha \frac{S_v}{N_H}  & -\alpha \frac{I_H}{N_H} - \mu  & 0\\
				0 & \alpha \frac{S_v}{N_H}  & \alpha \frac{I_H}{N_H} & - \mu\\
			\end{pmatrix}
		\end{equation}
		Then, we evaluate the Jacobian at the fixed point (or disease free equilibrium, DFE), yielding
		\begin{equation}
			J|_{DFE} = \begin{pmatrix}
				0& 0 & 0 & - \beta \\
				0  & -\gamma & 0 & \beta   \\
				0 & -\alpha \frac{C}{N_H}\frac{\delta}{\mu}  &  - \mu  & 0\\
				0 & \alpha \frac{C}{N_H}\frac{\delta}{\mu}  & 0 & - \mu\\
			\end{pmatrix}
		\end{equation}
		where $S_H=N_H$ has been considered.
		
		Finally, we obtain the eigenvalues of this matrix as
		\begin{equation}
			\begin{split}
				det(J|_{DFE} - \lambda \mathbb{I}) &= -\lambda \bigg[ - (\mu + \lambda)^2(\gamma+\lambda) +(\mu + \lambda) \beta\alpha \frac{C}{N_H}\frac{\delta}{\mu}  \bigg] = 0 \Rightarrow \\
				\lambda_0 &= 0 \\
				\lambda_\mu &= -\mu \\
				\lambda_{\pm} &= -\frac{(\gamma+\mu)}{2} \pm \frac{1}{2}\sqrt{ (\gamma-\mu)^2 +4\beta \alpha \frac{C}{N_H}\frac{\delta}{\mu} }
			\end{split}
		\end{equation}
	
		It is straightforward to see that all eigenvalues are real and the stability of the disease-free equilibrium is determined by the sign of the eigenvalues. $\lambda_\mu = -\mu <0$ as $\mu$ is defined positive, so  in order to discuss the stability of this fixed point, we need to study the $\lambda_{\pm}$ eigenvalues. $\lambda_{-}$ is always negative, but $\lambda_{+}$ changes sign depending on the values of the parameters. The threshold condition $\lambda_{+} = 0$ leads to:
		\begin{equation}
			\lambda_{+} = 0 \; \Rightarrow \; \frac{\beta \alpha}{\gamma \mu} \frac{C}{N_H}\frac{\delta}{\mu} = 1
			\label{eq:lambda+_general}
		\end{equation} 
		So, for $\displaystyle\frac{\beta \alpha}{\gamma \mu} \frac{C}{N_H}\frac{\delta}{\mu}<1 \, \Rightarrow \, \lambda_{+} < 0 $ the fixed point is stable and for $\displaystyle\frac{\beta \alpha}{\gamma \mu} \frac{C}{N_H}\frac{\delta}{\mu} >1 \, \Rightarrow \, \lambda_{+} > 0 $ a perturbation will grow in the direction of the eigenvector associated to $\lambda_{+}$. Thus, this threshold defines the basic reproduction number,
		\begin{equation}
			R_0 = \frac{\beta \alpha}{\gamma \mu}\frac{C}{N_H}\frac{\delta}{\mu}
		\end{equation}
		
		If instead of $S_H=N_H$ one considers any initial condition of hosts, $S_H(0)$, the basic reproduction number is given by,
		\begin{equation}
			R_0 = \frac{\beta \alpha}{\gamma \mu}\frac{C}{N_H}\frac{\delta}{\mu}\frac{S_H(0)}{N_H}
		\end{equation}
		
\subsection*{Next Generation Matrix method}
	
	The previous result can also be obtained by means of the NGM method, which is explained in detail in\citep{Diekmann2010}. Basically the method is based in decomposing the Jacobian in the form $\textbf{J}=\mathbf{T}+\mathbf{\Sigma}$, where $\mathbf{T}$ is the \textit{transmission part}, that describes the production of new infections, and $\mathbf \Sigma$ the \textit{transition part}, that describes changes of state (including death). Then, it can be proved \citep{Diekmann2010} that the \textit{basic reproduction number} $R_0$ is given by the spectral radius (i.e. the largest eigenvalue) of the (next generation) matrix $\mathbf{K}=- \mathbf{T} \mathbf{\Sigma}^{-1}$.
    \begin{equation}
			\mathbf K=- \mathbf T \mathbf \Sigma^{-1} = 
			\begin{pmatrix}
				\frac{\beta \alpha}{\gamma \mu}\frac{C}{N_H}\frac{\delta}{\mu} & \frac{\beta}{\mu} \\ \ ,
				0 & 0 \\
			\end{pmatrix}
			\label{eq:NGM_general}
		\end{equation}
		with,
		\begin{equation}
			\mathbf T = 
			\begin{pmatrix}
				0 & \beta\frac{N_H}{N_H} \\
				0 & 0
			\end{pmatrix} \qquad
			\mathbf \Sigma = 
			\begin{pmatrix}
				-\gamma & 0 \\
				\alpha \frac{C}{N_H}\frac{\delta}{\mu} & -\mu
			\end{pmatrix} \; \Rightarrow \;
			-{\mathbf\Sigma}^{-1} = 
			\begin{pmatrix}
				\frac{1}{\gamma}& 0 \\
				\frac{\alpha}{\gamma\mu}\frac{C}{N_H}\frac{\delta}{\mu} & \frac{1}{\mu}
			\end{pmatrix}
			\label{eq:F-V_general}
		\end{equation}
		
		The basic reproduction number is the spectral radius of this matrix so:
		\begin{equation*}
			det(\mathbf K-\sigma\mathbb{I}) = 0 \implies
			\begin{vmatrix}
				\frac{\beta \alpha}{\gamma \mu}\frac{C}{N_H}\frac{\delta}{\mu} -\sigma  &  \frac{\beta}{\mu } \\
				0 & -\sigma \\
			\end{vmatrix}
			= (-\sigma)\bigg(\frac{\beta \alpha}{\gamma \mu}\frac{C}{N_H}\frac{\delta}{\mu} -\sigma\bigg)  =0 \implies 
		\end{equation*}
		\begin{equation}
			\sigma = \frac{\beta \alpha}{\gamma \mu}\frac{C}{N_H}; \quad \sigma = 0
		\end{equation}
		Therefore, the basic reproduction number is
		\begin{equation}
			R_0 = \frac{\beta \alpha}{\gamma \mu}\frac{C}{N_H}\frac{\delta}{\mu}
		\end{equation}

        If instead of $S_H=N_H$ one considers any initial condition of hosts, $S_H(0)$, the basic reproduction number is given by,
		\begin{equation}
			R_0 = \frac{\beta \alpha}{\gamma \mu}\frac{C}{N_H}\frac{\delta}{\mu}\frac{S_H(0)}{N_H}
		\end{equation}

\section{Calculation of $R_0$ for non-stationary vector populations} \label{app:R0_non_stationary}

    A simple way of extending the definition of $R_0$ in the case of non-stationary vector populations consists of averaging the number of  secondary infections produced by an infected individual along one generation, that is equivalent to averaging the instantaneous definition of $R_0$, namely $R_0^i$, over one generation,
    \begin{equation}\label{eq:R_eff_to_integrate}
        \overline{R_0}=\avg{R_0^i(t)}\Big|\limitss{0}{t_g}=\frac{R_0}{N_v^*}\avg{N_v(t)}\Big|\limitss{0}{t_g}=\frac{R_0}{N_v^*}\frac{1}{t_g}\int_0^{t_g}N_v(t) \, \dif t \ ,
    \end{equation}
    where the integral in \cref{eq:R_eff_to_integrate} is solved as
    \begin{equation}
        \int_0^{t_g}N_v(t) \, \dif t=\claudator{N_v^*t -\frac{1}{\mu}\parentesi{N_v(0)-N_v^*}e^{-\mu t}}\limitss{0}{t_g}=N_v^*t_g -\frac{1}{\mu}\parentesi{N_v(0)-N_v^*}\claudator{e^{-\mu t_g} - 1} \ .
    \end{equation}
    Thus, the basic reproduction number for non-stationary vector populations is given by
    \begin{equation}
        \overline{R_0}=\frac{R_0}{N_v^*}\left\{N_v^*-\frac{1}{\mu t_g}\claudator{N_v(0)-N_v^*}\claudator{e^{-\mu t_g}-1}\right\} \ ,
        \label{eq:r0mdef}
    \end{equation}
    where the generation time, $t_g$, is \cref{eq:generationtime}.
    \cref{eq:r0mdef} can be rewritten as,
    \begin{equation}
        \overline{R_0}=\avg{R_{0}^{i}(t)}\Big|\limitss{0}{t_g}=R_0\claudator{1-\frac{1}{\tau}\parentesi{f-1}\parentesi{e^{-\tau}-1}}=R_0\cdot\mathcal{F} \ ,
       \label{eq:R0_non_stationary2}
    \end{equation}
    where $\tau=1+\mu/\gamma$ and $\mathcal{F}$ is the expression in brackets, which accounts for the effect of the decaying vector population on the stationary $R_0$.
    
    In our approach, a generation is defined as the time elapsed in the following sequence of processes: 1) A host individual becomes infected; 2) The infected host passes the disease to a susceptible vector; 3) The infected vector dies. Basically, the time elapsed from the first to the last process is the time in which new infections can be produced, i.e. $t_g$ \cref{eq:generationtime}.



\newpage

\begin{thebibliography}{36}
\expandafter\ifx\csname natexlab\endcsname\relax\def\natexlab#1{#1}\fi
\providecommand{\url}[1]{\texttt{#1}}
\providecommand{\href}[2]{#2}
\providecommand{\path}[1]{#1}
\providecommand{\DOIprefix}{doi:}
\providecommand{\ArXivprefix}{arXiv:}
\providecommand{\URLprefix}{URL: }
\providecommand{\Pubmedprefix}{pmid:}
\providecommand{\doi}[1]{\href{http://dx.doi.org/#1}{\path{#1}}}
\providecommand{\Pubmed}[1]{\href{pmid:#1}{\path{#1}}}
\providecommand{\bibinfo}[2]{#2}
\ifx\xfnm\relax \def\xfnm[#1]{\unskip,\space#1}\fi
\bibitem[{Anderson(1991)}]{Anderson1991}
\bibinfo{author}{Anderson, R.~M.} (\bibinfo{year}{1991}).
\newblock \bibinfo{title}{Discussion: The {Kermack}-{McKendrick} epidemic
  threshold theorem}.
\newblock {\it \bibinfo{journal}{Bulletin of Mathematical Biology}\/},  {\it
  \bibinfo{volume}{53}\/}, \bibinfo{pages}{3--32}.
  \DOIprefix\doi{10.1016/S0092-8240(05)80039-4}.
\bibitem[{Athni et~al.(2020)}]{Athni_2020}
\bibinfo{author}{Athni, T.} et~al. (\bibinfo{year}{2020}).
\newblock \bibinfo{title}{How vector-borne disease shaped the course of human
  history}.
\newblock \DOIprefix\doi{10.22541/au.159135318.82566392}.
\bibitem[{Baca{\"e}r(2007)}]{Bacaer2007}
\bibinfo{author}{Baca{\"e}r, N.} (\bibinfo{year}{2007}).
\newblock \bibinfo{title}{Approximation of the basic reproduction number r0 for
  vector-borne diseases with a periodic vector population}.
\newblock {\it \bibinfo{journal}{Bulletin of Mathematical Biology}\/},  {\it
  \bibinfo{volume}{69}\/}, \bibinfo{pages}{1067--1091}.
  \DOIprefix\doi{10.1007/s11538-006-9166-9}.
\bibitem[{Baca{\"e}r \& Gomes(2009)}]{Bacaer2009}
\bibinfo{author}{Baca{\"e}r, N.}, \& \bibinfo{author}{Gomes, M. G.~M.}
  (\bibinfo{year}{2009}).
\newblock \bibinfo{title}{On the final size of epidemics with seasonality}.
\newblock {\it \bibinfo{journal}{Bulletin of Mathematical Biology}\/},  {\it
  \bibinfo{volume}{71}\/}, \bibinfo{pages}{1954}.
  \DOIprefix\doi{10.1007/s11538-009-9433-7}.
\bibitem[{Baca{\"e}r \& Guernaoui(2006)}]{Bacaer2006}
\bibinfo{author}{Baca{\"e}r, N.}, \& \bibinfo{author}{Guernaoui, S.}
  (\bibinfo{year}{2006}).
\newblock \bibinfo{title}{The epidemic threshold of vector-borne diseases with
  seasonality}.
\newblock {\it \bibinfo{journal}{Journal of Mathematical Biology}\/},  {\it
  \bibinfo{volume}{53}\/}, \bibinfo{pages}{421--436}.
  \DOIprefix\doi{10.1007/s00285-006-0015-0}.
\bibitem[{Bragard et~al.(2013)Bragard, Caciagli, Lemaire, Lopez-Moya,
  MacFarlane, Peters, Susi \& Torrance}]{Bragard2013}
\bibinfo{author}{Bragard, C.}, \bibinfo{author}{Caciagli, P.},
  \bibinfo{author}{Lemaire, O.}, \bibinfo{author}{Lopez-Moya, J.},
  \bibinfo{author}{MacFarlane, S.}, \bibinfo{author}{Peters, D.},
  \bibinfo{author}{Susi, P.}, \& \bibinfo{author}{Torrance, L.}
  (\bibinfo{year}{2013}).
\newblock \bibinfo{title}{Status and prospects of plant virus control through
  interference with vector transmission}.
\newblock {\it \bibinfo{journal}{Annual Review of Phytopathology}\/},  {\it
  \bibinfo{volume}{51}\/}, \bibinfo{pages}{177--201}.
  \DOIprefix\doi{10.1146/annurev-phyto-082712-102346}.
\bibitem[{Brauer(2008)}]{Brauer2008}
\bibinfo{author}{Brauer, F.} (\bibinfo{year}{2008}).
\newblock \bibinfo{title}{Compartmental models in epidemiology}.
\newblock In \bibinfo{editor}{F.~Brauer}, \bibinfo{editor}{P.~van~den
  Driessche}, \& \bibinfo{editor}{J.~Wu} (Eds.), {\it
  \bibinfo{booktitle}{Mathematical Epidemiology}\/} (pp.
  \bibinfo{pages}{19--79}).
\newblock \bibinfo{address}{Berlin, Heidelberg}: \bibinfo{publisher}{Springer
  Berlin Heidelberg}.
\newblock \DOIprefix\doi{10.1007/978-3-540-78911-6_2}.
\bibitem[{Brauer et~al.(2016)Brauer, Castillo-Chavez, Mubayi \&
  Towers}]{Brauer2016}
\bibinfo{author}{Brauer, F.}, \bibinfo{author}{Castillo-Chavez, C.},
  \bibinfo{author}{Mubayi, A.}, \& \bibinfo{author}{Towers, S.}
  (\bibinfo{year}{2016}).
\newblock \bibinfo{title}{Some models for epidemics of vector-transmitted
  diseases}.
\newblock {\it \bibinfo{journal}{Infectious Disease Modelling}\/},  {\it
  \bibinfo{volume}{1}\/}, \bibinfo{pages}{79--87}.
  \DOIprefix\doi{10.1016/j.idm.2016.08.001}.
\bibitem[{Castillo-Chavez \& Thieme(1995)}]{Thieme1995}
\bibinfo{author}{Castillo-Chavez, C.}, \& \bibinfo{author}{Thieme, H.~R.}
  (\bibinfo{year}{1995}).
\newblock \bibinfo{title}{Asymptotically autonomous epidemic models}.
\newblock In \bibinfo{editor}{D.~Arino, O.~Axelrod},
  \bibinfo{editor}{M.~Kimmel}, \& \bibinfo{editor}{M.~Langlais} (Eds.), {\it
  \bibinfo{booktitle}{{Mathematical Population Dynamics: Analysis of
  Heterogeneity, Vol. I, Theory of Epidemics}}\/} (pp.
  \bibinfo{pages}{33--50}).
\newblock \bibinfo{publisher}{Wuerz publishing (Winnipeg)}.
\newblock \URLprefix
  \url{https://ecommons.cornell.edu/bitstream/handle/1813/31834/BU-1248-M.pdf}.
\bibitem[{Chowell(2017)}]{Chowel2017}
\bibinfo{author}{Chowell, G.} (\bibinfo{year}{2017}).
\newblock \bibinfo{title}{Fitting dynamic models to epidemic outbreaks with
  quantified uncertainty: A primer for parameter uncertainty, identifiability,
  and forecasts}.
\newblock {\it \bibinfo{journal}{Infectious Disease Modelling}\/},  {\it
  \bibinfo{volume}{2}\/}, \bibinfo{pages}{379--398}.
  \DOIprefix\doi{10.1016/j.idm.2017.08.001}.
\bibitem[{Diekmann \& Heesterbeek(2000)}]{Diekmann2000}
\bibinfo{author}{Diekmann, O.}, \& \bibinfo{author}{Heesterbeek, J. A.~P.}
  (\bibinfo{year}{2000}).
\newblock {\it \bibinfo{title}{Mathematical Epidemiology of Infectious
  Diseases. Model Building, Analysis and Interpretation}\/}.
\newblock \bibinfo{address}{Chichester (UK)}: \bibinfo{publisher}{John Wiley
  and Sons}.
\bibitem[{Diekmann et~al.(2010)Diekmann, Heesterbeek \& Roberts}]{Diekmann2010}
\bibinfo{author}{Diekmann, O.}, \bibinfo{author}{Heesterbeek, J. A.~P.}, \&
  \bibinfo{author}{Roberts, M.~G.} (\bibinfo{year}{2010}).
\newblock \bibinfo{title}{The construction of next-generation matrices for
  compartmental epidemic models}.
\newblock {\it \bibinfo{journal}{Journal of The Royal Society Interface}\/},
  {\it \bibinfo{volume}{7}\/}, \bibinfo{pages}{873--885}.
  \DOIprefix\doi{10.1098/rsif.2009.0386}.
\bibitem[{Esteva \& Vargas(1998)}]{Esteva1998}
\bibinfo{author}{Esteva, L.}, \& \bibinfo{author}{Vargas, C.}
  (\bibinfo{year}{1998}).
\newblock \bibinfo{title}{Analysis of a dengue disease transmission model}.
\newblock {\it \bibinfo{journal}{Mathematical Biosciences}\/},  {\it
  \bibinfo{volume}{150}\/}, \bibinfo{pages}{131--151}.
  \DOIprefix\doi{10.1016/S0025-5564(98)10003-2}.
\bibitem[{Garms et~al.(1979)Garms, Walsh \& Davies}]{garms1979studies}
\bibinfo{author}{Garms, R.}, \bibinfo{author}{Walsh, J.}, \&
  \bibinfo{author}{Davies, J.} (\bibinfo{year}{1979}).
\newblock \bibinfo{title}{Studies on the reinvasion of the onchocerciasis
  control programme in the volta river basin by simulium damnosum si with
  emphasis on the south-western areas.}
\newblock {\it \bibinfo{journal}{Tropenmedizin und Parasitologie}\/},  {\it
  \bibinfo{volume}{30}\/}, \bibinfo{pages}{345--362}.
\bibitem[{Giménez-Romero et~al.(2021)Giménez-Romero, Grau, Hendriks \&
  Matias}]{Gimenez2021}
\bibinfo{author}{Giménez-Romero, A.}, \bibinfo{author}{Grau, A.},
  \bibinfo{author}{Hendriks, I.~E.}, \& \bibinfo{author}{Matias, M.~A.}
  (\bibinfo{year}{2021}).
\newblock \bibinfo{title}{Modelling parasite-produced marine diseases: The case
  of the mass mortality event of pinna nobilis}.
\newblock {\it \bibinfo{journal}{Ecological Modelling}\/},  {\it
  \bibinfo{volume}{459}\/}, \bibinfo{pages}{109705}.
  \DOIprefix\doi{10.1016/j.ecolmodel.2021.109705}.
\bibitem[{Huang et~al.(2020)Huang, Reyes-Caldas, Mann, Seifbarghi, Kahn,
  Almeida, Béven, Heck, Hogenhout \& Coaker}]{HUANG20201379}
\bibinfo{author}{Huang, W.}, \bibinfo{author}{Reyes-Caldas, P.},
  \bibinfo{author}{Mann, M.}, \bibinfo{author}{Seifbarghi, S.},
  \bibinfo{author}{Kahn, A.}, \bibinfo{author}{Almeida, R.~P.},
  \bibinfo{author}{Béven, L.}, \bibinfo{author}{Heck, M.},
  \bibinfo{author}{Hogenhout, S.~A.}, \& \bibinfo{author}{Coaker, G.}
  (\bibinfo{year}{2020}).
\newblock \bibinfo{title}{Bacterial vector-borne plant diseases: Unanswered
  questions and future directions}.
\newblock {\it \bibinfo{journal}{Molecular Plant}\/},  {\it
  \bibinfo{volume}{13}\/}, \bibinfo{pages}{1379--1393}.
  \DOIprefix\doi{10.1016/j.molp.2020.08.010}.
\bibitem[{Kamgang \& Sallet(2008)}]{Kamgang2008}
\bibinfo{author}{Kamgang, J.~C.}, \& \bibinfo{author}{Sallet, G.}
  (\bibinfo{year}{2008}).
\newblock \bibinfo{title}{Computation of threshold conditions for
  epidemiological models and global stability of the disease-free equilibrium
  (dfe)}.
\newblock {\it \bibinfo{journal}{Mathematical Biosciences}\/},  {\it
  \bibinfo{volume}{213}\/}, \bibinfo{pages}{1--12}.
  \DOIprefix\doi{10.1016/j.mbs.2008.02.005}.
\bibitem[{Kao \& Eisenberg(2018)}]{Kao2018}
\bibinfo{author}{Kao, Y.-H.}, \& \bibinfo{author}{Eisenberg, M.~C.}
  (\bibinfo{year}{2018}).
\newblock \bibinfo{title}{Practical unidentifiability of a simple vector-borne
  disease model: Implications for parameter estimation and intervention
  assessment}.
\newblock {\it \bibinfo{journal}{Epidemics}\/},  {\it \bibinfo{volume}{25}\/},
  \bibinfo{pages}{89--100}. \DOIprefix\doi{10.1016/j.epidem.2018.05.010}.
\bibitem[{Kermack \& McKendrick(1927)}]{Kermack1927}
\bibinfo{author}{Kermack, W.~O.}, \& \bibinfo{author}{McKendrick, A.~G.}
  (\bibinfo{year}{1927}).
\newblock \bibinfo{title}{A contribution to the mathematical theory of
  epidemics}.
\newblock {\it \bibinfo{journal}{Proceedings of the Royal Society of London.
  Series A}\/},  {\it \bibinfo{volume}{115}\/}, \bibinfo{pages}{700--721}.
  \DOIprefix\doi{10.1098/rspa.1927.0118}.
\bibitem[{Lashari \& Zaman(2011)}]{Lashari2011}
\bibinfo{author}{Lashari, A.~A.}, \& \bibinfo{author}{Zaman, G.}
  (\bibinfo{year}{2011}).
\newblock \bibinfo{title}{Global dynamics of vector-borne diseases with
  horizontal transmission in host population}.
\newblock {\it \bibinfo{journal}{Computers \& Mathematics with
  Applications}\/},  {\it \bibinfo{volume}{61}\/}, \bibinfo{pages}{745--754}.
  \DOIprefix\doi{10.1016/j.camwa.2010.12.018}.
\bibitem[{Laukó(2006)}]{Lauko2006}
\bibinfo{author}{Laukó, I.~G.} (\bibinfo{year}{2006}).
\newblock \bibinfo{title}{Stability of disease free sets in epidemic models}.
\newblock {\it \bibinfo{journal}{Mathematical and Computer Modelling}\/},  {\it
  \bibinfo{volume}{43}\/}, \bibinfo{pages}{1357--1366}.
  \DOIprefix\doi{10.1016/j.mcm.2005.06.011}.
\bibitem[{Macdonald(1957)}]{Macdonald1957}
\bibinfo{author}{Macdonald, G.} (\bibinfo{year}{1957}).
\newblock {\it \bibinfo{title}{The epidemiology and control of malaria}\/}.
\newblock \bibinfo{address}{Oxford (UK)}: \bibinfo{publisher}{Oxford University
  Press}.
\bibitem[{Martcheva(2015)}]{MartchevaBook}
\bibinfo{author}{Martcheva, M.} (\bibinfo{year}{2015}).
\newblock {\it \bibinfo{title}{An Introduction to Mathematical
  Epidemiology}\/}.
\newblock \bibinfo{address}{New York}: \bibinfo{publisher}{Springer}.
\newblock \DOIprefix\doi{10.1007/978-1-4899-7612-3}.
\bibitem[{Rockl{\"o}v \& Dubrow(2020)}]{Rocklov2020}
\bibinfo{author}{Rockl{\"o}v, J.}, \& \bibinfo{author}{Dubrow, R.}
  (\bibinfo{year}{2020}).
\newblock \bibinfo{title}{Climate change: an enduring challenge for
  vector-borne disease prevention and control}.
\newblock {\it \bibinfo{journal}{Nature Immunology}\/},  {\it
  \bibinfo{volume}{21}\/}, \bibinfo{pages}{479--483}.
  \DOIprefix\doi{10.1038/s41590-020-0648-y}.
\bibitem[{Roosa \& Chowell(2019)}]{Roosa2019}
\bibinfo{author}{Roosa, K.}, \& \bibinfo{author}{Chowell, G.}
  (\bibinfo{year}{2019}).
\newblock \bibinfo{title}{Assessing parameter identifiability in compartmental
  dynamic models using a computational approach: application to infectious
  disease transmission models}.
\newblock {\it \bibinfo{journal}{Theoretical Biology and Medical Modelling}\/},
   {\it \bibinfo{volume}{16}\/}, \bibinfo{pages}{1}.
  \DOIprefix\doi{10.1186/s12976-018-0097-6}.
\bibitem[{Rybicki(2015)}]{Rybicki2015}
\bibinfo{author}{Rybicki, E.~P.} (\bibinfo{year}{2015}).
\newblock \bibinfo{title}{A top ten list for economically important plant
  viruses}.
\newblock {\it \bibinfo{journal}{Archives of Virology}\/},  {\it
  \bibinfo{volume}{160}\/}, \bibinfo{pages}{17--20}.
  \DOIprefix\doi{10.1007/s00705-014-2295-9}.
\bibitem[{Schneider et~al.(2020)Schneider, van~der Werf, Cendoya, Mourits,
  Navas-Cort{\'e}s, Vicent \& Oude~Lansink}]{Schneider2020}
\bibinfo{author}{Schneider, K.}, \bibinfo{author}{van~der Werf, W.},
  \bibinfo{author}{Cendoya, M.}, \bibinfo{author}{Mourits, M.},
  \bibinfo{author}{Navas-Cort{\'e}s, J.~A.}, \bibinfo{author}{Vicent, A.}, \&
  \bibinfo{author}{Oude~Lansink, A.} (\bibinfo{year}{2020}).
\newblock \bibinfo{title}{Impact of xylella fastidiosa subspecies pauca in
  european olives}.
\newblock {\it \bibinfo{journal}{Proceedings of the National Academy of
  Sciences}\/},  {\it \bibinfo{volume}{117}\/}, \bibinfo{pages}{9250--9259}.
  \DOIprefix\doi{10.1073/pnas.1912206117}.
\bibitem[{Schumacher \& Campbell(2018)}]{SCHUMACHER2018352}
\bibinfo{author}{Schumacher, S.~K.}, \& \bibinfo{author}{Campbell, J.~I.}
  (\bibinfo{year}{2018}).
\newblock \bibinfo{title}{Chapter 56 - travel medicine}.
\newblock In \bibinfo{editor}{R.~P. Olympia}, \bibinfo{editor}{R.~M.
  O’Neill}, \& \bibinfo{editor}{M.~L. Silvis} (Eds.), {\it
  \bibinfo{booktitle}{Urgent Care Medicine Secrets}\/} (pp.
  \bibinfo{pages}{352--357}).
\newblock \bibinfo{publisher}{Elsevier}.
\newblock \DOIprefix\doi{10.1016/B978-0-323-46215-0.00056-2}.
\bibitem[{Shah \& Jyoti(2013)}]{Shah2013}
\bibinfo{author}{Shah, N.}, \& \bibinfo{author}{Jyoti, G.}
  (\bibinfo{year}{2013}).
\newblock \bibinfo{title}{{SEIR} model and simulation for vector borne
  diseases}.
\newblock {\it \bibinfo{journal}{Applied Mathematics}\/},  {\it
  \bibinfo{volume}{4}\/}, \bibinfo{pages}{13 -- 17}.
  \DOIprefix\doi{10.4236/am.2013.48A003}.
\bibitem[{Thieme(1992)}]{Thieme1992}
\bibinfo{author}{Thieme, H.~R.} (\bibinfo{year}{1992}).
\newblock \bibinfo{title}{Convergence results and a {Poincaré-Bendixson}
  trichotomy for asymptotically autonomous differential equations}.
\newblock {\it \bibinfo{journal}{Journal of Mathematical Biology}\/},  {\it
  \bibinfo{volume}{30}\/}, \bibinfo{pages}{755--763}.
  \DOIprefix\doi{10.1007/BF00173267}.
\bibitem[{Tumber et~al.(2014)Tumber, Alston, Fuller et~al.}]{tumber2014pierce}
\bibinfo{author}{Tumber, K.}, \bibinfo{author}{Alston, J.},
  \bibinfo{author}{Fuller, K.} et~al. (\bibinfo{year}{2014}).
\newblock \bibinfo{title}{Pierce's disease costs {California} \$104 million per
  year}.
\newblock {\it \bibinfo{journal}{California Agriculture}\/},  {\it
  \bibinfo{volume}{68}\/}, \bibinfo{pages}{20--29}.
\bibitem[{{van den Driessche}(2017)}]{VandenDriessche2017}
\bibinfo{author}{{van den Driessche}, P.} (\bibinfo{year}{2017}).
\newblock \bibinfo{title}{Reproduction numbers of infectious disease models}.
\newblock {\it \bibinfo{journal}{Infectious Disease Modelling}\/},  {\it
  \bibinfo{volume}{2}\/}, \bibinfo{pages}{288--303}.
  \DOIprefix\doi{10.1016/j.idm.2017.06.002}.
\bibitem[{Wei et~al.(2008)Wei, Li \& Martcheva}]{Martcheva2008}
\bibinfo{author}{Wei, H.-M.}, \bibinfo{author}{Li, X.-Z.}, \&
  \bibinfo{author}{Martcheva, M.} (\bibinfo{year}{2008}).
\newblock \bibinfo{title}{An epidemic model of a vector-borne disease with direct transmission and time delay}.
\newblock {\it \bibinfo{journal}{Journal of Mathematical Analysis and
  Applications}\/},  {\it \bibinfo{volume}{342}\/}, \bibinfo{pages}{895--908}.
  \DOIprefix\doi{10.1016/j.jmaa.2007.12.058}.
\bibitem[{Wesley \& Allen(2009)}]{Wesley2009}
\bibinfo{author}{Wesley, C.~L.}, \& \bibinfo{author}{Allen, L.~J.}
  (\bibinfo{year}{2009}).
\newblock \bibinfo{title}{The basic reproduction number in epidemic models with
  periodic demographics}.
\newblock {\it \bibinfo{journal}{Journal of Biological Dynamics}\/},  {\it
  \bibinfo{volume}{3}\/}, \bibinfo{pages}{116--129}.
  \DOIprefix\doi{10.1080/17513750802304893}.
\newblock \bibinfo{note}{PMID: 22880824}.
\bibitem[{WHO()}]{WHO}
WHO (\bibinfo{year}{2018}).
\newblock \bibinfo{title}{Fact sheets of vector-borne diseases}.
\newblock \URLprefix
  \url{https://www.hiobs.org/lib/file/manager/Vector-Borne_Disease_General_Fact_Sheet.pdf}.
\bibitem[{Zhao et~al.(2020)Zhao, Musa, Hebert, Cao, Ran, Meng, He \&
  Qin}]{Zhao2020}
\bibinfo{author}{Zhao, S.}, \bibinfo{author}{Musa, S.~S.},
  \bibinfo{author}{Hebert, J.~T.}, \bibinfo{author}{Cao, P.},
  \bibinfo{author}{Ran, J.}, \bibinfo{author}{Meng, J.}, \bibinfo{author}{He,
  D.}, \& \bibinfo{author}{Qin, J.} (\bibinfo{year}{2020}).
\newblock \bibinfo{title}{Modelling the effective reproduction number of
  vector-borne diseases: the yellow fever outbreak in {Luanda, Angola}
  2015-2016 as an example}.
\newblock {\it \bibinfo{journal}{PeerJ}\/},  {\it \bibinfo{volume}{8}\/},
  \bibinfo{pages}{e8601--e8601}. \DOIprefix\doi{10.7717/peerj.8601}.

\end{thebibliography}
\end{document}